\documentclass[aps,physrev,preprint,titlepage]{revtex4-2}

\usepackage[utf8]{inputenc}
\usepackage{amsmath}
\usepackage{amssymb}
\usepackage{amsfonts}
\usepackage{xcolor}
\usepackage[margin = 2.5cm]{geometry}
\usepackage{graphicx}
\usepackage{ragged2e}
\usepackage{caption}
\DeclareCaptionJustification{justified}{\justifying}
\usepackage{subcaption}
\usepackage{esint}
\usepackage{tikz}
\usepackage{physics}
\usepackage{array}
\newcolumntype{I}{!{\vrule width 1.2pt}}
\usepackage[hyperfootnotes = true, colorlinks = true, linkcolor = blue, citecolor = purple]{hyperref}
\usetikzlibrary{shapes.misc,decorations.markings,arrows,decorations.pathmorphing,backgrounds,matrix,patterns,positioning,decorations.pathreplacing,calc}

\graphicspath{{./figures/}}

\definecolor{dgreen}{rgb}{0,0.5,0}
\definecolor{dpink}{rgb}{1,0.3,0.3}
\definecolor{darkblue}{rgb}{0,0,0.6}
\definecolor{purple}{rgb}{0.4,.2,0.7}
\definecolor{lblue}{rgb}{0.5,0.5,1.0}
\definecolor{ggray}{rgb}{0.85,.85,0.85}
\definecolor{darkgreen}{rgb}{0,0.5,0}

\def\tr{\mathrm{tr}}

\newcommand{\bes}{\begin{equation} \begin{split} }
\newcommand{\ees}{\end{split} \end{equation} }

\renewcommand{\d}{\mathrm{d}}

\begin{document}
\title{Regge trajectories from the adjoint sector\texorpdfstring{\\}{ } of Matrix Quantum Mechanics}

\author{Igor R. Klebanov}
\author{Henry W. Lin}
\author{Pavel Meshcheriakov}
\affiliation{Joseph Henry Laboratories, Princeton University, Princeton, NJ 08544, USA}

\date{\today}

\begin{abstract}
We reexamine the large $N$ limit of SU$(N)$ symmetric quantum mechanics of a Hermitian matrix whose singlet sector is well known to be exactly solvable via free fermions. When the Fermi level approaches a maximum of the potential, there is critical behavior corresponding to string theory in two dimensions.  We uncover new phenomena in the adjoint sector by solving the Marchesini-Onofri equation both numerically and analytically using semiclassical approximations: at criticality, the spectrum is governed by Regge trajectories with energy eigenvalues growing according to $\Delta^2 \sim n/ \alpha'$. In the dual 2D string theory, we interpret these states as oscillatory excitations of a ``short'' folded open string.  Up to sub-leading corrections, this Regge behavior is essentially universal and is insensitive to the particular potential we choose to approach criticality. Slightly away from criticality, the highly excited states transition into ``long strings'' that extend far into the Liouville direction. 
\end{abstract}

\maketitle
\section{Introduction and Summary}
The generalization of QCD from the SU$(3)$ gauge group to SU$(N)$ has been an important theoretical method for studying the dynamics of strong interactions. Taking the 't Hooft large $N$ limit, where $g_{\rm YM}^2 N$ is held fixed, results in a restriction to the planar Feynman diagrams \cite{tHooft:1973alw}. While this produces some important simplifications, this has not led to a solution of QCD. However, soon after the original idea \cite{tHooft:1973alw}, Brezin, Itzykson, Parisi, and Zuber \cite{Brezin:1977sv} discovered that there are simpler models with $N\times N$ matrix degrees of freedom whose large $N$ limits are exactly solvable.
One of the richest such models, which we will revisit in this paper, is the SU$(N)$ symmetric quantum mechanics of a Hermitian matrix $X$ with potential $V(X)$. This model is often called Matrix Quantum Mechanics (MQM). Its SU$(N)$ invariant sector is separable because the $N$ eigenvalues of $X$ act as non-interacting fermions moving in the potential $V$ \cite{Brezin:1977sv}. In particular, the ground state is obtained by filling the $N$ lowest energy levels, so the Fermi energy $\mu_F$ equals the $N$th single-particle energy.  

The dynamics of states transforming in non-trivial representations of SU$(N)$ is more complicated, but luckily Marchesini and Onofri \cite{Marchesini:1979yq} have derived an elegant integral equation describing the large $N$ limit of states in the adjoint representation. For the case of the quartic potential \cite{Brezin:1977sv}, 
\begin{equation}\label{eq:quartpot}
    V_{\rm quartic}(X)=\tr\left(\frac{1}{2}X^2+gX^4\right)\ ,
\end{equation}
with $\hbar=\frac{1}{N}$, the numerical solutions of the Marchesini-Onofri (MO) equation were first obtained in \cite{Marchesini:1979yq,Casartelli:1979iz}. For positive $g$, as the excitation number $n$ increases, the eigenvalues were found to rapidly approach the WKB formula
\begin{equation}\label{eq:WKB}
     \Delta_n^{\rm high} = (n+1)\omega(g)+\eta(g)+ \mathcal{O}\left(1/n\right)\ .
\end{equation} 
Here, $\omega(g)$ coincides with the singlet energy gap \eqref{frequency}, while $\eta(g)$ can be expressed as an integral \eqref{MO-potential} \cite{Marchesini:1979yq,Casartelli:1979iz}.   

As discovered in \cite{Brezin:1977sv}, the large $N$ limit of MQM with the quartic potential is stable not only for positive $g$, but also for negative $g$ above the critical value $g_c=-\frac{\sqrt 2}{6\pi}$ (this corrects the expression in \cite{Brezin:1977sv}).
As $g$ approaches $g_c$ from above, the Fermi level $\mu_F$ approaches the maximum of the potential from below, i.e. $V_{\rm max}- \mu_F \equiv \mu \rightarrow 0$. One can equivalently think of parameterizing the potential in terms of $\mu$ instead of $g$; in this limit, $g-g_c\sim \mu\log\mu$.
In the double scaling limit where $N\mu$ is held fixed, the contributions of large random surfaces of all topologies are included \cite{Brezin:1989ss,Gross:1990ay,Ginsparg:1990as}. This limit of MQM describes 2D quantum gravity coupled to a massless scalar field, so it is often called the $c=1$ matrix model. Since the dynamics of 2D quantum gravity is described by the Liouville field theory \cite{Polyakov:1981rd}, the double scaling limit of MQM is dual to string theory in {\it two} dimensions, where the Liouville dimension $\phi$ is encoded in the matrix eigenvalues \cite{Das:1990kaa,Gross:1990st,Sengupta:1990bt}. 
The potential \eqref{eq:quartpot} corresponds to the string scale $\alpha'=\frac{1}{2}$. 
A review of developments in 2D string theory up to 1991 can be found in \cite{Klebanov:1991qa}.

Both the ground state of the $c=1$ matrix model and its low-lying excitations are described by the solvable theory of free fermions coming from the singlet sector of SU$(N)$. On the other hand, the thermal partition function, $\tr\, e^{-2\pi R H}$, which corresponds to a circular imaginary time with radius $R$, also receives contributions from all the allowed non-singlet sectors, starting with the adjoint. However, it was found \cite{Gross:1990ub} that the contribution of the singlet sector alone exhibits the $R\rightarrow \frac{\alpha'}{R}$ duality expected in the continuum formulation of the compact boson coupled to the Liouville theory. This led the authors of \cite{Gross:1990ub,Gross:1990md} to conjecture that the contributions of the non-singlet sectors implement the effects of the Berezinskii-Kosterlitz-Thouless (BKT) vortices on the discretized random surfaces. Further support for this picture was provided in \cite{Boulatov:1991xz,Boulatov:1991fp}, and a possible application of the non-singlets to 2D black holes was proposed in \cite{Kazakov:2000pm}.

The simplest non-trivial representation, the adjoint, corresponds to the vortex-antivortex pair \cite{Gross:1990md,Boulatov:1991fp}. In terms of the $c=1$ matrix model with real time, the states in the adjoint representation correspond to the motion of a folded open string \cite{Maldacena:2005hi}. We can view this as a special case of inserting an adjoint Wilson line (or more generally, inserting adjoint operators connected by a Wilson line) in a holographic gauge theory, which creates a folded string that comes in from the UV \cite{Maldacena:1998im, Rey:1998ik, Klebanov:2006jj, Maldacena:2018vsr}.  
The regime where the string becomes very long has been studied in the approximation of the upside down harmonic oscillator potential \cite{Maldacena:2005hi,Fidkowski:2005ck,Karczmarek:2008sc, Betzios:2017yms, Balthazar:2018qdv,Betzios:2022pji, Ahmadain:2022gfw}. 

The folded open string picture of the adjoint sector suggests that the eigenvalues of the $\mu\rightarrow 0$ limit of the MO equation should have a clear string theoretic interpretation. In this paper, we carry out a comprehensive study of this limit in the MO equation without making any approximations in the potential. Recent work using two-point function bootstrap \cite{Cho:2025vws} (see also \cite{Cho:2024kxn, Adams:2025nww}) found that the few lowest eigenvalues $\Delta_n$ are in excellent agreement with the numerical solution of the MO equation for small $\mu$. In particular, the gap $\Delta_1$ between the adjoint and singlet ground states approaches $\approx 0.7416$ \cite{Cho:2025vws}. The fact that $\Delta_1$  does not grow logarithmically negates the conjecture in \cite{Gross:1990md,Boulatov:1991fp,Boulatov:1991xz}. Nevertheless, the finiteness of the adjoint gap implies that the low energy dynamics is dominated by the singlet sector,
since the singlet excitation energy $\omega$ vanishes as
${\pi \sqrt 2}/\log \frac{4}{|g_c|\mu}$ near criticality.

In this paper, we go further and carry out a study of many energy eigenvalues of the MO equation in the quartic model, finding a sequence of states with eigenvalues labeled with a positive integer $n$,
\begin{equation}\label{eq:regge-phi4}
    \Delta_n^{\rm Regge}\approx \sqrt{2 n+\frac{2}{3}}- \frac{2\sqrt 2}{\pi}\ ,\qquad n\lesssim  n_\text{max}\approx\frac{1}{4\pi^2}\log^2\mu\ .
\end{equation}
Although technically this formula was only derived for $n\gg 1$, we find that even for $n=1$ the agreement with the numerical results is excellent.
We will interpret these states
as the result of oscillations of the tip of a folded open string.
The $\sim\sqrt n$ growth of the rest energies of states is characteristic of Regge trajectories in string theory.

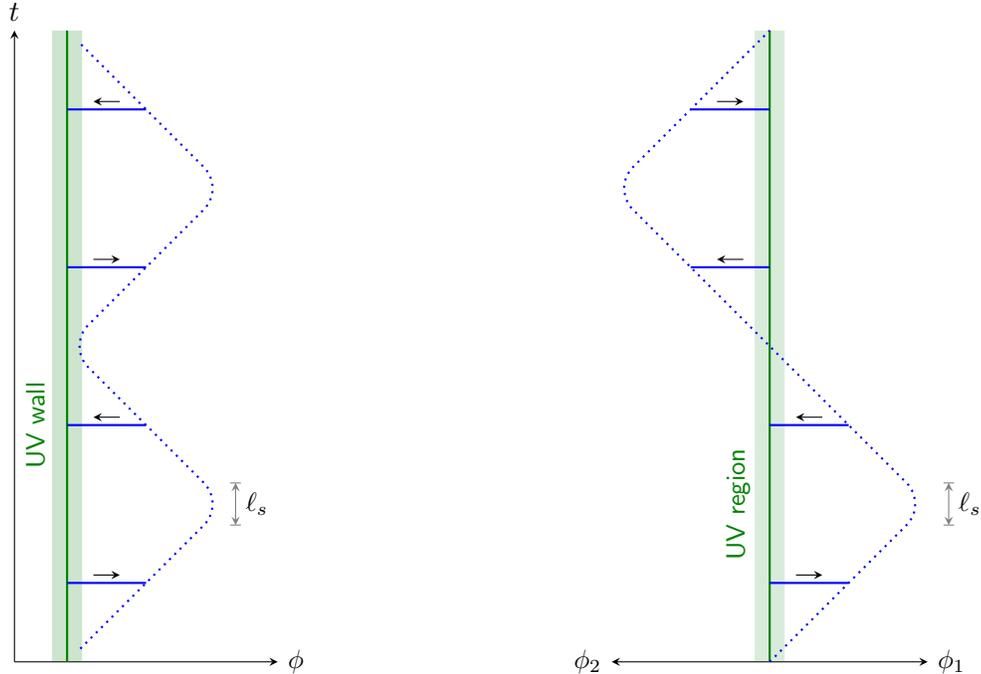
\begin{figure}[ht]
    \begin{subfigure}[b]{0.49\textwidth}
        \centering
        \begin{tikzpicture}[>=stealth, scale=0.7]
            \draw[->, >=stealth] (-1, -4.5) -- (4, -4.5) node[right] {\footnotesize $\phi$};
            \draw[->, >=stealth] (-1, -4.5) -- (-1, 7.5) node[above] {$t$};

            \draw[dgreen, line width=0.4cm, opacity=0.2] (0, -4.5) -- (0, 7.5);
            \draw[thick, dgreen] (0, -4.5) -- (0, 7.5);
            \node[dgreen, rotate=90] at (-0.6, 0) {\footnotesize \sf{UV wall}};

            \draw[dotted, blue, thick, rounded corners=0.4cm] (0.25, -4.25) -- (3, -1.5) -- (0, 1.5) -- (3, 4.5) -- (0.25, 7.25);
            
            \draw[<->, >=stealth, gray, thin] (3.2, -1.9) -- (3.2, -1.1) node[midway, right, black] {\footnotesize $\ell_s$};
            \draw[gray, thin] (3.1, -1.9) -- (3.3, -1.9);
            \draw[gray, thin] (3.1, -1.1) -- (3.3, -1.1);

            \draw[blue, thick] (0, -3.0) -- (1.5, -3.0); \draw[->, black] (0.5, -2.85) -- (1.0, -2.85);
            \draw[blue, thick] (0, 0.0) -- (1.5, 0.0); \draw[<-, black] (0.5, 0.15) -- (1.0, 0.15);
            \draw[blue, thick] (0, 3.0) -- (1.5, 3.0); \draw[->, black] (0.5, 3.15) -- (1.0, 3.15);
            \draw[blue, thick] (0, 6.0) -- (1.5, 6.0); \draw[<-, black] (0.5, 6.15) -- (1.0, 6.15);
        \end{tikzpicture}
    \end{subfigure}
    \begin{subfigure}[b]{0.49\textwidth}
        \begin{tikzpicture}[>=stealth, scale=0.7]
            \draw[->, >=stealth] (0, -4.5) -- (3., -4.5) node[right] {\footnotesize $\phi_1$};
            \draw[->, >=stealth] (0, -4.5) -- (-3., -4.5) node[left] {\footnotesize $\phi_2$};
            \draw[dgreen, line width=0.4cm, opacity=0.15] (0, -4.5) -- (0, 7.5);
            \draw[thick, dgreen] (0, -4.5) -- (0, 7.5);
            \node[dgreen, rotate=90] at (-0.6, -1.5) {\footnotesize \sf{UV region}};

            \draw[dotted, blue, thick, rounded corners=0.4cm] (0, -4.5) -- (3, -1.5) -- (0, 1.5) -- (-3, 4.5) -- (0, 7.5);
            \draw[<->, >=stealth, gray, thin] (3.4, -1.9) -- (3.4, -1.1) node[midway, right, black] {\footnotesize $\ell_s$};
            \draw[gray, thin] (3.3, -1.9) -- (3.5, -1.9);
            \draw[gray, thin] (3.3, -1.1) -- (3.5, -1.1);
            \draw[blue, thick] (0, -3.0) -- (1.5, -3.0); \draw[->, black] (0.5, -2.85) -- (1.0, -2.85);
            \draw[blue, thick] (0, 0.0) -- (1.5, 0.0); \draw[<-, black] (0.5, 0.15) -- (1.0, 0.15);
            \draw[blue, thick] (0, 3.0) -- (-1.5, 3.0); \draw[->, black] (-0.5, 3.15) -- (-1.0, 3.15);
            \draw[blue, thick] (0, 6.0) -- (-1.5, 6.0); \draw[<-, black] (-0.5, 6.15) -- (-1.0, 6.15);
        \end{tikzpicture}
    \end{subfigure}
    \caption{{\it Left}: A cartoon of the oscillatory folded string motion. The string extends from the ``UV wall'' at $\phi=0$ 
    to some $\phi_\text{tip}$ (the dotted {\color{blue} blue} line). As long as the tip reaches sufficiently far, 
    the dynamics of the string is universal. For most of the trajectory the tip of the string travels nearly at the speed of light, but near the turning point it has nonzero acceleration for a few string lengths. {\it Right}: Two Liouville regions attached at $\phi=0$, and the $\mathbb{Z}_2$ symmetry acts as $\phi\rightarrow -\phi$. The string endpoints are fixed at $\phi=0$, while the fold oscillates between the two regions. }
    \label{fig:strings}
\end{figure}

For the quartic potential, there is a $\mathbb{Z}_2$ reflection symmetry, under which the wave functions pick up factor $(-1)^n$. We interpret states with odd and even $n$ as two separate Regge trajectories. In considering their interpretation in terms of the 2D string theory, we suggest that there are two Liouville regions glued together at $\phi=0$ and related by reflection.
As shown in figure \ref{fig:strings}, the endpoints of the string are fixed at the $\phi=0$ interface, while the tip oscillates between the regions of positive and negative $\phi$. We may apply a $\mathbb{Z}_2$ orbifold projection, after which only the Regge trajectory with even $n$ remains.

\begin{figure}[ht]
    \centering
    \includegraphics[width=.47\columnwidth]{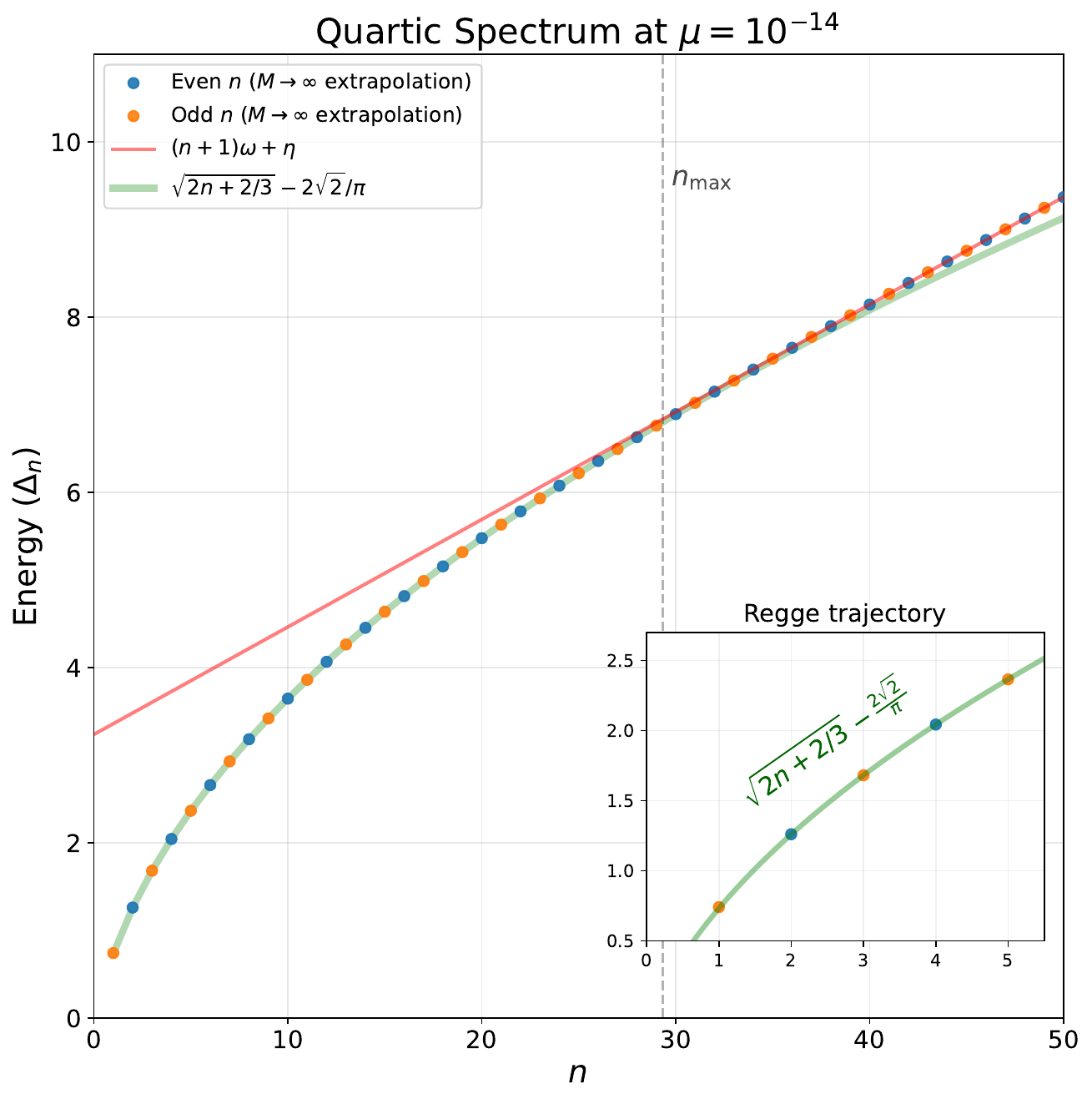}\hspace{0.5cm}
    \includegraphics[width=.47\columnwidth]{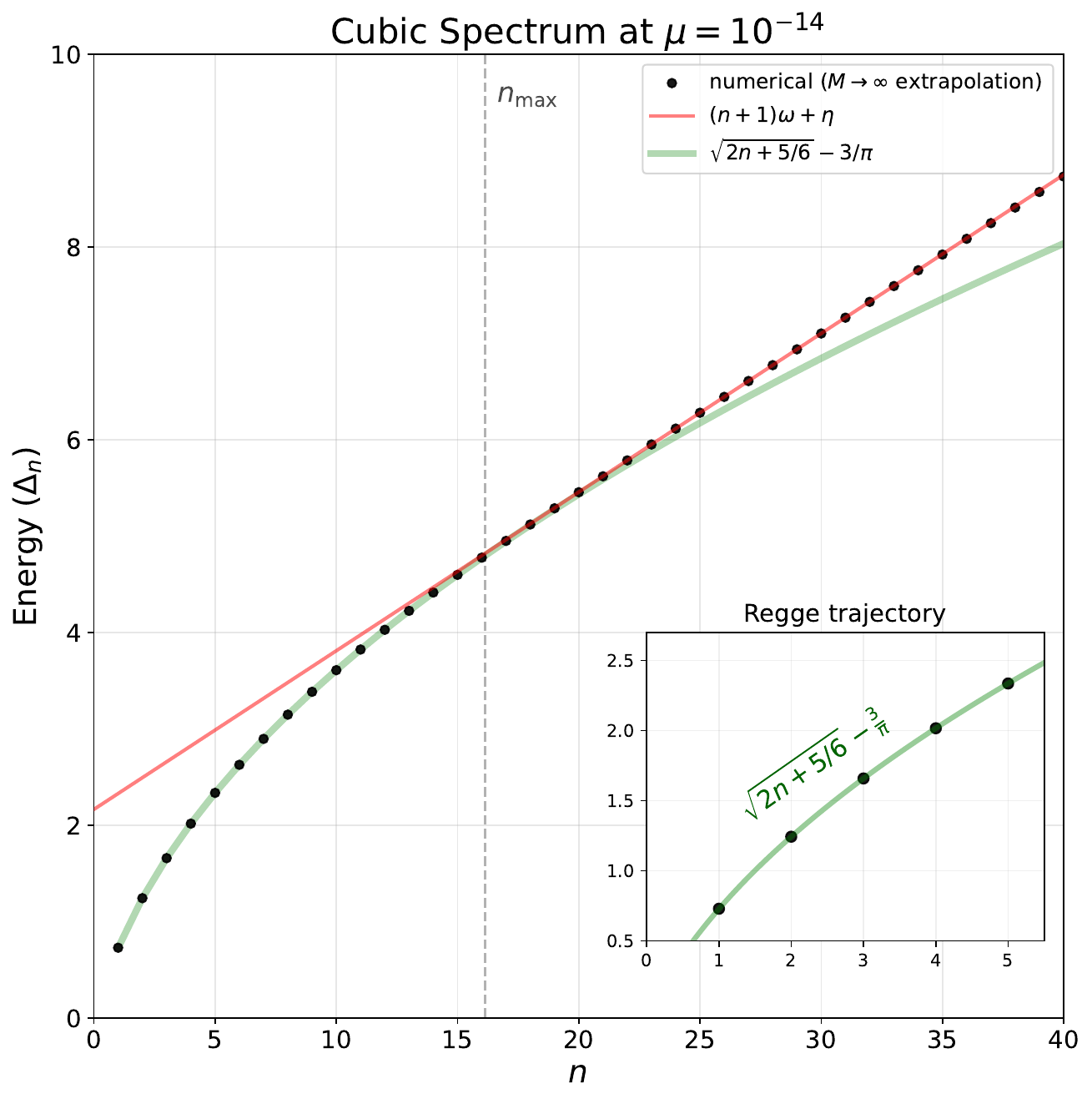}
    \caption{{\it Left}: the MO spectrum $\Delta_n$ for the quartic theory near criticality $ g\to g_c$. The $\mathbb{Z}_2$ even (odd) states are shown with {\color{blue}blue} ({\color{orange}orange}) dots. Here $n_{\rm max}$ is given in \eqref{eq:nmax-phi4}, the approximate value where the {\color{red} red} curve (high energy WKB) touches the {\color{dgreen} green} curve (the Regge trajectory). In the inset panels, we show the first few MO eigenvalues, which closely follow the Regge formula \eqref{eq:regge-phi4}. The deviation at small $n$, which is nearly imperceptible, is presumably due to corrections to the Regge formula and non-universal features of the ``UV wall.'' {\it Right}: spectrum for the cubic theory near criticality. It is qualitatively similar but there is no $\mathbb{Z}_2$ symmetry. The transition value $n_\text{max}$ between the Regge behavior and the high-energy WKB is given by \eqref{eq:nmax-phi3}.
    }
    \label{fig:spectrum}
\end{figure}

\begin{figure}
    \centering
    \includegraphics[width=\linewidth]{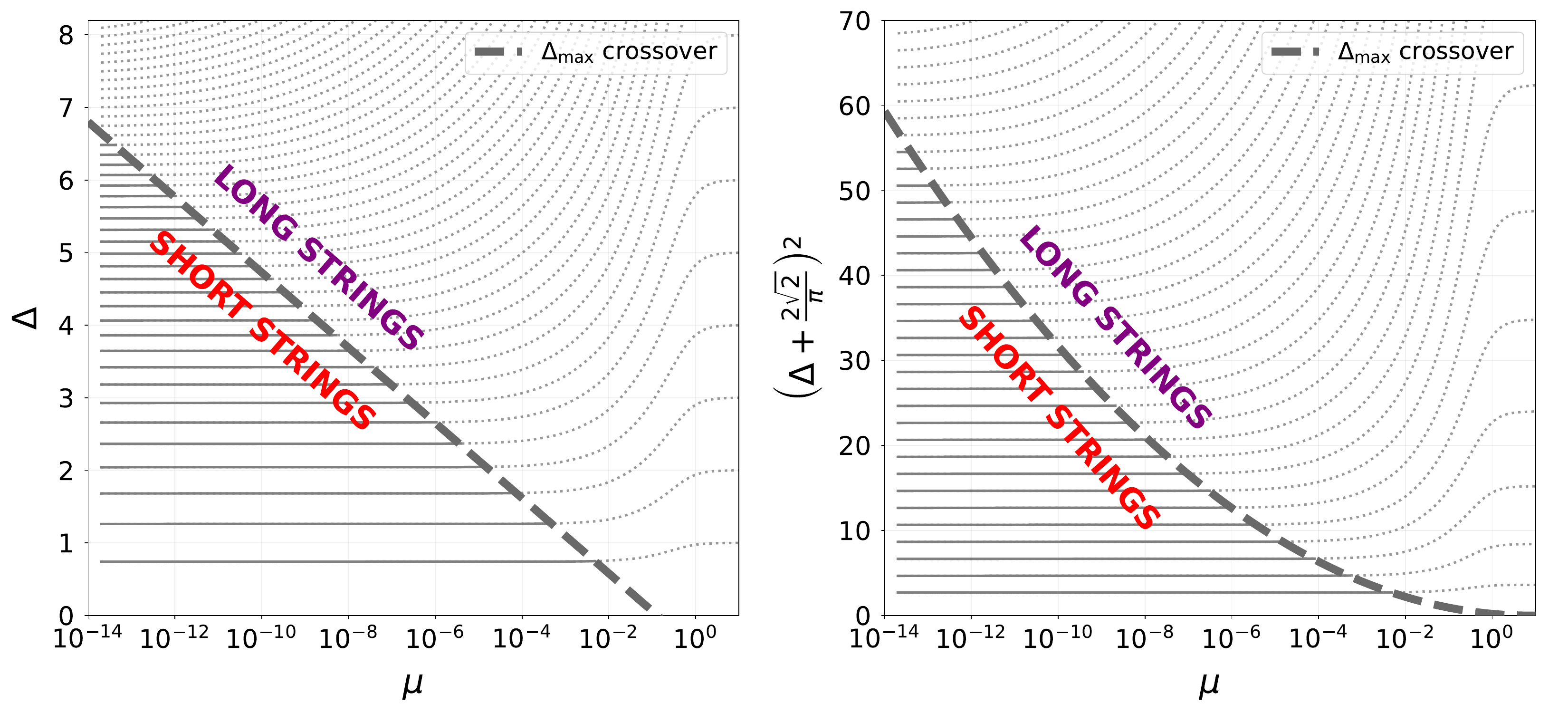}
    \caption{Eigenvalues from diagonalizing the discretized MO equation for quartic theory with $M=36000$. On the right panel, we see that $\left(\Delta+\tfrac{2\sqrt{2}}{\pi}\right)^2$ becomes evenly spaced for small $\mu$, which is the Regge behavior $\Delta^2 \sim n$. (We have included the constant offset in \eqref{eq:regge-phi4} so that the spacings look more even for small $n$). On the left panel, we see that at large $\mu$, $\Delta$ becomes evenly spaced as expected from a weakly coupled theory ($g\to 0$). At energies much bigger than the crossover energy $\Delta \gg \Delta_\text{max}$, we expect the $\Delta$'s to be evenly spaced but with a much finer spacing \eqref{eq:WKB} than the spacings in the weak coupling theory. Note that $\mu$ needs to be quite small $\mu \lesssim 10^{-4}$ in order to resolve several levels of the Regge trajectory. The transition between long and short strings takes place approximately at $\Delta_{\rm max}$, with the corresponding $n_{\rm max}$ given in \eqref{eq:nmax-phi4}.}
    \label{fig:mu}
\end{figure}

For $n\gtrsim n_{\rm max}$, the WKB formula \eqref{eq:WKB} becomes applicable, and $\eta(\mu)\approx\frac{1}{2\sqrt2\pi}\log\frac{4}{|g_c|\mu}-\frac{2\sqrt2}{\pi}$.
A similar approximation was derived in \cite{Gross:1990md}, where the quartic potential corresponding to $\alpha'=\frac{1}{4}$ was used.
We confirm the approximations \eqref{eq:WKB} and \eqref{eq:regge-phi4} using high-precision numerical solutions of the MO equation with tiny values of $\mu$, such as $\mu=10^{-14}$. As shown in figures \ref{fig:spectrum} and \ref{fig:mu}, our numerical solutions interpolate very smoothly between the approximations $\Delta_n^{\rm Regge}$ for $n\lesssim n_{\rm max}$ and $\Delta_n^{\rm high}$ for $n\gtrsim n_{\rm max}$.

In addition to the quartic potential \eqref{eq:quartpot},
we have solved the MO equation for the near-critical cubic and double-well potentials.
For the cubic potential \eqref{cubicpot}, 
as the coupling is tuned to the critical value so that the triangulated surfaces become large, the Fermi level approaches the local maximum of the potential, $\mu\rightarrow 0$. While the MO equation is somewhat more complicated than in the quartic case, we find that the Regge and linear regimes are present again for $\mu\rightarrow 0$, and the interpolation between them is smooth.    
Since the cubic potential has no $\mathbb{Z}_2$ symmetry,
we find a clear correspondence with the 2D string theory containing the Liouville potential $\sim \mu e^{2\phi/\sqrt{\alpha'}}$. 
As shown on the left of figure \ref{fig:strings}, the endpoints of the string are attached to the boundary of the Liouville coordinate $\phi$, and the oscillations are made possible by the tip of the string reflecting from the boundary (an analogous oscillating folded string with the ends at the boundary of AdS space was studied in \cite{Klebanov:2006jj}).

The double well matrix potential near criticality 
describes the 2D type 0B string theory
\cite{Douglas:2003up,Takayanagi:2003sm}. In this case, the theory is stable for either sign of $\mu$. When we take the limits $|\mu|\rightarrow 0$, the qualitative features of the spectra are similar to those found for the quartic and cubic potentials.


\section{Marchesini-Onofri equation near criticality}

For the SU$(N)$ invariant (singlet) sector of MQM, it is well known \cite{Brezin:1977sv} that the system is described by $N$ non-interacting fermions. The ground state is obtained by filling the first $N$ energy levels in potential $V$. 
In the WKB approximation, the singlet excitation gap is given by the classical frequency $\omega(g)$ 
\begin{equation}\label{frequency}
     \frac{2\pi}{\omega(g)}=\frac{1}{\pi}\oint\limits\frac{\d x}{\rho(x)}, \qquad \rho(x) = \frac{1}{\pi}\sqrt{2(\mu_F-V(x))}\ .\\
\end{equation}
Here $\rho$ is the eigenvalue density (or the fermion density \cite{Brezin:1977sv}) in the ground state, which is supported between $x_1$ and $x_2$, the classical turning points of the potential $V(x)$ (for even potentials $x_2=-x_1=a$).
For the first non-trivial representation of SU$(N)$, which is the adjoint representation, the situation is more complicated. In \cite{Marchesini:1979yq}, it was shown that the spectrum of the adjoint sector, in the large-$N$ limit, is structured as follows: for each singlet state with energy $E_{s}^{\rm singlet}$ whose excitation does not modify the leading planar fermion density, there is an infinite tower of bound states in the adjoint sector with energies $E_{(s,n)}^{\rm adjoint}$
\begin{equation}
    E_{(s,n)}^{\rm adjoint}(g)=E_s^{\rm singlet}(g)+\Delta_n(g)\ ,\qquad n=1,2,\dots \ .
\end{equation}
At planar order, the structure of this tower does not depend on the choice of the $s$-th singlet state within this class: $\Delta_n(g)$ is a universal gap between the $n$-th state in the tower and the corresponding singlet state. For different macroscopic singlet backgrounds, the gaps become background dependent; here we focus on the standard ground state background.

To study these gaps in the large $N$ limit, Marchesini and Onofri \cite{Marchesini:1979yq} derived the singular integral equation
\begin{equation}\label{MO-eq}
    \Delta_n\, \Phi_n(x)=\fint_{x_1}^{x_2}\limits \d y\, \rho(y) 
    \frac{\Phi_n(x)-\Phi_n(y)}{(x-y)^2}\ ,
\end{equation}
where $\Phi_n(x)$ are the eigenfunctions associated with the corresponding adjoint states.

The MO equation \eqref{MO-eq} resembles the well-known 't Hooft equation \cite{THOOFT1974461}, which describes mesons in two-dimensional large $N$ QCD. In fact, for $V(x)=0$, the MO equation becomes equivalent to the 't Hooft equation for massless quarks, and $\Delta$ is then proportional to the meson mass-squared. In this case, the spectrum consists of two Regge trajectories of mesons \cite{Fateev:2009jf,Litvinov:2024riz,Artemev:2025cev}. This is qualitatively similar to the two Regge trajectories \eqref{eq:regge-phi4}.
For the 't Hooft model the trajectories are asymptotically linear, and their string theoretic interpretation is in terms of longitudinal oscillations of 2D open strings with moving endpoints \cite{Bardeen:1975gx}. 

The eigenfunctions $\Phi_n$, $n=1, 2, \ldots$, are orthonormal with respect to the measure $\rho(x) \d x$, e.g.,  
 $\int_{x_1}^{x_2}\limits \d x\, \rho(x) \, \Phi^*_n(x)\Phi_m(x)=\delta_{nm}\, $
while the adjoint representation constraint \cite{Marchesini:1979yq,Maldacena:2005hi}
    $\int_{x_1}^{x_2}\limits \d x\, \rho(x)  \, \Phi_n(x)=0 $
removes the constant eigenfunction with eigenvalue $0$. The lowest eigenvalue $\Delta_1$ corresponds to $\Phi_1(x)$, which is an odd function for the quartic potential \eqref{eq:quartpot}.

\begin{figure}
    \includegraphics[width=0.49\columnwidth]{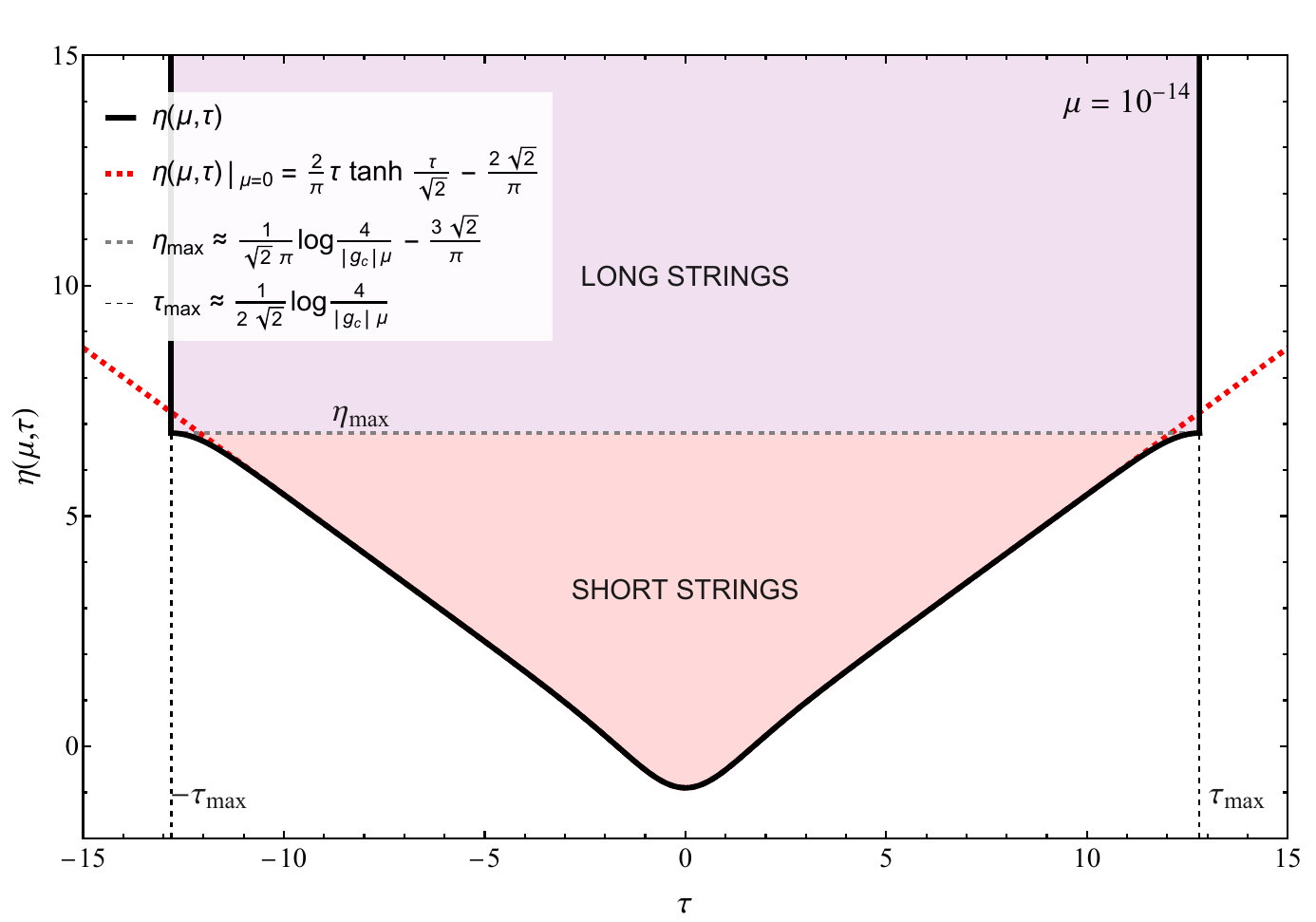}
    \includegraphics[width=.49\columnwidth]{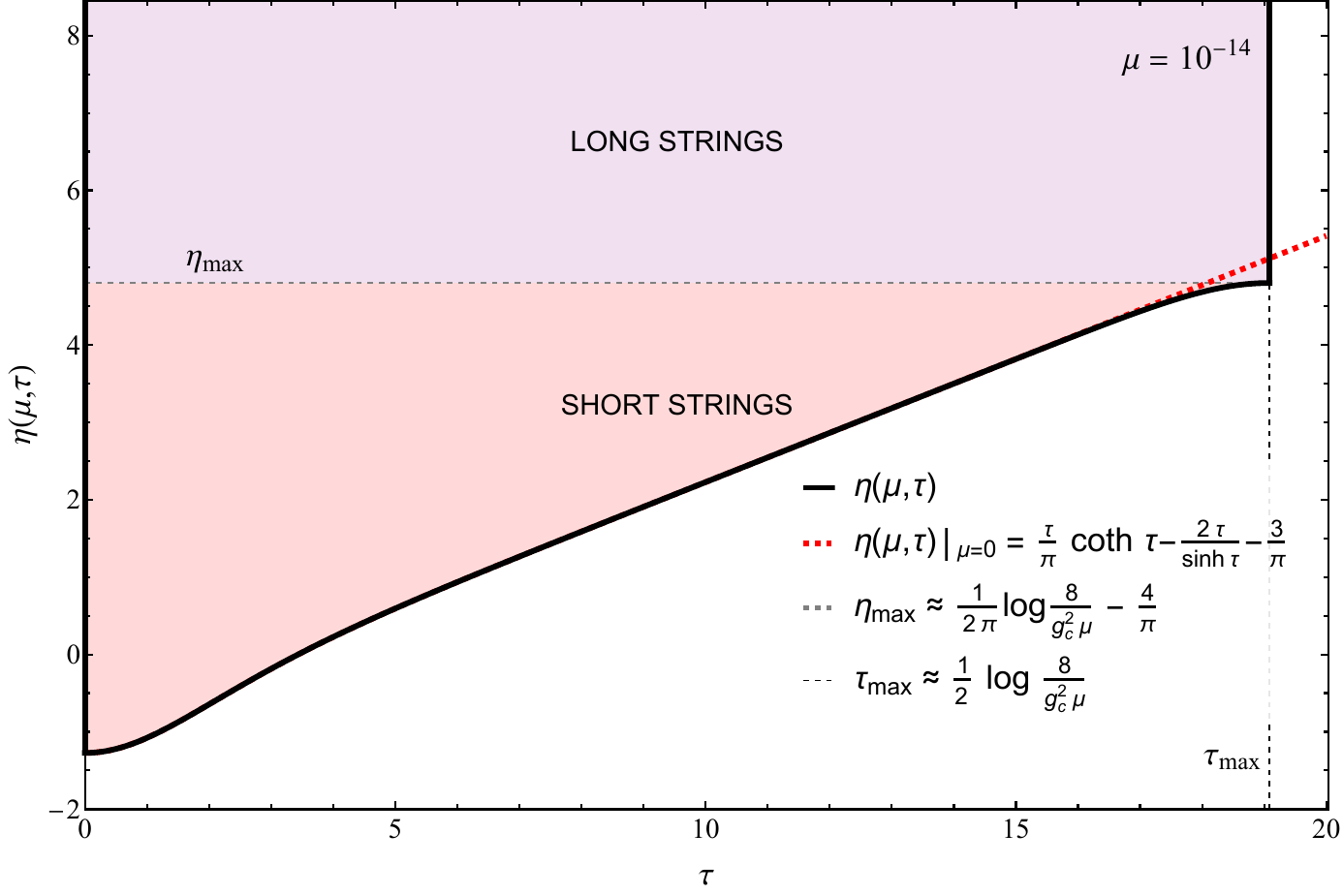}
    \caption{{\it Left:} the effective potential $\eta$ in the MO equation \eqref{MO-1} for the quartic potential \eqref{eq:quartpot}, using the time of flight variable $\tau$. We use $\mu=10^{-14}$, which shows a good separation between the UV region near $\tau=0$ and the two ``Liouville walls" at $\tau=\pm \tau_{\rm max}$. {\it Right}: the MO potential for the cubic using the time-of-flight variable $\tau$. At $\tau = 0$ there is the UV wall; at $\tau = \tau_{\rm max}$ there is the Liouville wall.}
    \label{fig:MO-potential}
\end{figure}

After introducing $\phi_n(x)=\rho(x) \,\Phi_n(x)$, the MO equation becomes
$\Delta_n\,\phi_n(x)=\mathcal{H}\,\phi_n(x)$. The effective Hamiltonian $\mathcal{H}$ is defined by \cite{Marchesini:1979yq}
\begin{align}\label{MO-1}
    \mathcal{H}\,\phi_n(x)=-\rho(x)\fint_{x_1}^{x_2}\limits \d y\,\frac{\phi_n(y)}{(x-y)^2} 
    + \eta(g,x)\phi_n(x)
    \ ,
\end{align}
where $\eta(g,x)$ defined below plays the role of ``potential,''
\begin{equation}\label{MO-potential}
    \eta(g,x)\equiv\fint_{x_1}^{x_2}\limits \d y\,\frac{\rho(y)}{(x-y)^2}\ , \qquad \eta(g)=\frac{\omega(g)}{\pi}\int_{x_1}^{x_2}\limits\frac{\d x}{\pi}\frac{\eta(g,x)}{\rho(x)}\ ,
\end{equation}
and the quantity $\eta(g)$ in \eqref{eq:WKB} corresponds to the average potential over one period. 
To study the MO equation,
we will find it convenient to switch from $x$ to the time of flight variable $\tau$ of the classical trajectory at energy $\mu_F$:
\begin{equation}\label{eq:time-of-flight}
    \tau = \frac{1}{\pi} \int^x\limits \frac{\d y }{\rho(y)}\ .
\end{equation}
This change of variables has proven useful for studying the singlet sector of MQM \cite{Das:1990kaa,Klebanov:1991qa}.
For the quartic potential, $\tau\in[-\tau_{\rm max}(g),\,\tau_{\rm max}(g)]$, and $\tau_{\rm max}$ diverges at criticality as $\frac{1}{2\sqrt 2}\log \frac{4}{|g_c|\mu}$. The classical frequency \eqref{frequency} for the quartic theory in terms of the new coordinate is $\omega(\mu)=\frac{\pi}{2\tau_{\rm max}(\mu)}$.  
The potential $\eta(g,\tau)$ at the critical and near critical point for quartic and cubic theories is shown in figure \ref{fig:MO-potential}. 

In order to focus on the Regge states of quartic theory \eqref{eq:regge-phi4}, let us study the MO equation at $g=g_c$. The change of variables is $x= \frac{1}{2\sqrt{|g_c|}} \tanh\frac{\tau}{\sqrt{2}}$, 
and
$\tau\in (-\infty,\infty)$.
The effective Hamiltonian nicely separates:
\begin{equation}\label{decoupledH-phi4}
    \mathcal{H}_{\rm quartic}(p,\tau)=p\coth\frac{\pi p}{\sqrt{2}}+\frac{2}{\pi}\left(\tau\tanh\frac{\tau}{\sqrt{2}}-\sqrt{2}\right)\ .
\end{equation}
Note that in this critical regime, the kinetic term reproduces, up to a constant factor in the definition of momentum and energy, the kinetic term in the 't Hooft model \cite{NARAYANAN200576}, \cite{Fateev:2009jf,Litvinov:2024riz,Artemev:2025cev}.

When studying sufficiently excited states, $\mathcal{H}$ may be approximated by
\begin{equation}\label{eq:ultrarel}
    \mathcal{H}_{\rm quartic}(p,\tau)\approx |p|+\frac{2}{\pi}\left(|\tau|-\sqrt{2}\right)\ , 
\end{equation}
which describes a massless particle moving in a linear potential. In the dual description of critical MQM in terms of 2D string theory, this may be explained by the motion of the tip of a folded string \cite{Bardeen:1975gx,Maldacena:2005hi}. We will refer to such strings as {\it short strings}. These short strings can be long compared to the string length $\ell_s \sim \sqrt{\alpha'}$ but are short enough that they do not feel the Liouville potential.

The simplest setup, which is relevant to MQM with the quartic potential \eqref{eq:quartpot}, involves oscillations along a spatial Liouville coordinate $\phi \in (-\infty, \infty)$, with the endpoints fixed at $\phi=0$ (see figure \ref{fig:strings}).
This string has one fold, which acts as a massless particle \cite{Bardeen:1975gx,Maldacena:2005hi}, so that the energy of the string with the fold located at $\phi$ is
\begin{equation}
\label{eq:foldedstring}
    E= |p_\phi| + \frac{1}{\pi \alpha'} |\phi| + \text{const}\ ,
\end{equation}
where we used the fact that the string tension is $\frac{1}{2\pi\alpha'}$ and included a factor of 2 for two string segments. For $\alpha'=\frac{1}{2}$, this agrees with \eqref{eq:ultrarel} upon the identification $\phi=\tau$.
Applying the Bohr-Sommerfeld quantization
    $\oint p_\phi\, \d\phi = 2\pi n$,
where the integral is over the cycle where $\phi$ begins at
$-\pi\alpha' E$, turns around at $\pi\alpha' E$ and returns to the original position, we find
$2\pi \alpha' E_n^2= 2\pi n$. Therefore, $E_n= \sqrt{\frac{n}{\alpha'}}$, which agrees with the leading term in \eqref{eq:regge-phi4}. One can also solve for the eigenfunctions of this Hamiltonian exactly in terms of Fresnel integrals, for large energies this essentially gives Bohr-Sommerfeld quantization with a Maslov index $\gamma=\frac{1}{2}$, which leads to the improved formula \eqref{eq:regge-phi4} for the $\mu=0$ Hamiltonian \eqref{decoupledH-phi4}. Note that even for the first bound state $\Delta_1$, it has a relative error $\sim 1\%$, which rapidly improves with increasing excitation number (see Appendix \ref{app:numerics}). 

The presence of a linear dilaton background in the 2D string theory endows the fold with an effective mass proportional to the slope $Q=\frac{2}{\sqrt{\alpha'}}$ \cite{Maldacena:2005hi}. Indeed, expanding the exact kinetic term in \eqref{decoupledH-phi4} at low momenta gives $\frac{p^2}{2M_{\rm fold}}$ with $M_{\rm fold}=\frac{3}{2\pi\sqrt {\alpha'}}=\frac{3Q}{4\pi}$. Yet, the kinetic term differs considerably from the standard relativistic expression, since the correction to the ultrarelativistic limit $|p|$ is $\sim |p|\exp(-2\pi \sqrt{\alpha'} |p|).$ 
A qualitatively similar exponential correction is found in 2D string theory with the linear dilaton (see Appendix \ref{app:Dilaton-back}), a more quantitative match is left for future work.

A natural question is whether our results depend sensitively on the UV. Since the continuum string is emerging from the discrete sum over planar diagrams, we may ask whether the Regge trajectory depends on whether we use a triangular random lattice, etc. This is equivalent to asking whether our results depend on the matrix potential $V(X)$. 
We find that the results are essentially independent of the choice of matrix potential, up to sub-leading corrections in $n$ and $\log |\mu|$. 
For the cubic potential 
\begin{equation}\label{cubicpot}
    V_{\rm cubic}(X)=\tr\left(\frac{1}{2}X^2+gX^3\right)\ ,
\end{equation}
which generates triangulated random surfaces embedded in one dimension \cite{Brezin:1989ss,Klebanov:1991qa},
the large $N$ limit is stable for $g^2 \leq \frac{1}{15 \pi}$. Near the critical coupling, it corresponds to 2D string theory with $1/\alpha'$ equal to the curvature of the potential at the local maximum; hence, $\alpha'=1$.
The absence of $\mathbb{Z}_2$ symmetry makes us expect a single Regge trajectory.

To  study the MO equation for \eqref{cubicpot} at the critical coupling $g_c=-\frac{1}{\sqrt{15 \pi}}$, it is again convenient to change to the time-of-flight variable \eqref{eq:time-of-flight}: $x=\frac{1}{6g_c}\left (1-3\tanh^2\frac{\tau}{2}\right)$, and
now $\tau$ runs from $0$ to $\infty$. The critical MO Hamiltonian is
\begin{equation}\label{decoupledH-phi3}
    \mathcal{H}_{\rm cubic}(p,\tau)=
    p\coth(\pi p\tanh{\frac{\tau}{2}})+
    \frac{\tau}{\pi}\coth\tau - \frac{2\tau}{\sinh\tau}-\frac{3}{\pi}\ .
\end{equation}
Unlike in 
\eqref{decoupledH-phi4}, for the cubic model the kinetic term is not a function of $p$ alone. However, for large $|p|$, it is again proportional to $|p|$ up to exponentially small corrections. Using the semiclassical quantization of $\mathcal{H}_{\rm cubic}$ (with Maslov index $\gamma=\frac{1}{2}$), we find the eigenvalues:
\begin{equation}\label{eq:regge-phi3}
    \Delta^{\rm cubic}_n \approx \sqrt{2n+\frac{5}{6}}-\frac{3}{\pi}\ .  
\end{equation}
This approximation provides good accuracy even for $n=1$, where the numerical solution of the MO equation \eqref{MO-eq} gives $\Delta^{\rm cubic}_1\approx 0.7298$ and the relative error with \eqref{eq:regge-phi3} is $\sim 0.2\%$ (see Appendix \ref{app:numerics}). Near the critical point, the approximation (\ref{eq:regge-phi3})
is valid for $n\lesssim n^{\rm cubic}_{\rm max}$, where 
\begin{equation}
\label{nmaxcubic}
    n^{\rm cubic}_{\rm max}\approx\frac{1}{8\pi^2}\log^2\mu\ .
\end{equation}
The discussion of the semiclassical folded string for the cubic potential proceeds analogously to the quartic, but now the complete cycle consists of the fold starting at 
$\pi\alpha' E$ reflecting from the ``UV wall" at $\phi=0$, and returning to the original position (see the left of figure \ref{fig:strings}).
In this case, the semiclassical quantization gives $(E_n^{\rm cubic})^2 =\frac{2n}{\alpha'} + \cdots $, which agrees with the leading term in \eqref{eq:regge-phi3}.

The double-well potential $V_{\rm dw}(X)=-\frac{1}{2}X^2+gX^4$ describes type 0B string theory in two dimensions with $\alpha'=\frac{1}{2}$ \cite{Douglas:2003up,Takayanagi:2003sm}. Here, the critical coupling is $g_c= \frac{1}{3\pi}$, and for $g<g_c$, $\mu>0$ (in the singlet sector, the two wells are filled with the same Fermi level) the semiclassical spectrum is
\begin{equation}\label{eq:regge-dw-plus}
\Delta^{\rm dw^+}_n\approx
    \sqrt{2n+\frac{1}{3}}-\frac{2}{\pi},\quad n\lesssim n^{\rm dw^+}_{\rm max}\approx\frac{1}{8\pi^2}\log^2\mu\ .
\end{equation}
For the double-well potential, it is also possible to have $\mu<0$ \cite{Klebanov:1991qa,Douglas:2003up}; this corresponds to $g>g_c$
\begin{equation}\label{eq:regge-dw}
\Delta^{\rm dw^-}_n\approx
    \sqrt{2n+\frac{2}{3}}-\frac{4}{\pi},\quad n\lesssim n^{\rm dw^-}_{\rm max}\approx\frac{1}{2\pi^2}\log^2|\mu|\ .
\end{equation}
The $\mathbb{Z}_2$ symmetry of the double-well potential in the case $\mu<0$ distinguishes the R-R and NS-NS sectors \cite{Douglas:2003up,Takayanagi:2003sm}, so the eigenvalues with odd and even $n$ form two separate Regge trajectories.

To see the two Regge trajectories analytically (i.e., to see the alternating terms in the expression for energy levels \eqref{eq:regge-phi4}, \eqref{eq:regge-dw}), it is necessary to use more sophisticated analytical methods, one of which is the integrability-based Fateev-Lukyanov-Zamolodchikov (FLZ) method developed for the 't Hooft model \cite{Fateev:2009jf,Litvinov:2024riz,Artemev:2025cev}. Note that our analytical formulas reproduce the spectrum of short strings with very high accuracy (see table \ref{tab:numerics}). We expect corrections to our formulas to be exponentially small in contrast
to the larger corrections to the WKB eigenvalues of the 't Hooft equation \cite{Fateev:2009jf,Litvinov:2024riz,Artemev:2025cev}.

We now turn to the highly excited levels (long strings) $n\gtrsim n_{\rm max}$, which correspond to states exploring the full $\tau$ interval up to the endpoints $\tau=\pm\tau_{\rm max}(\mu)$. In contrast to the short-string sector, where the relevant motion is confined well inside the interval, the long-string states are sensitive to the fact that $\eta(\mu,\tau)$ terminates at finite $|\tau|=\tau_{\rm max}(\mu)$ with value $\eta_{\rm max} = \eta(\mu,\tau_\text{max})$ and effectively imposes hard walls. Consequently, the semiclassical turning points are pinned at $\tau=\pm\tau_{\rm max}(\mu)$ for the quartic potential (or $\tau=0, \tau_{\rm max}(\mu)$ for cubic), and the Bohr--Sommerfeld quantization (with Maslov index $\gamma=1$) becomes that of a particle in a finite box with a slowly varying potential in the interior. This immediately leads to a linear growth of the spectrum with $n$, in agreement with the long-string WKB law \eqref{eq:WKB}.

For the quartic theory \eqref{eq:quartpot} the quantities $\omega (g)$ \eqref{frequency} and $\eta (g)$ \eqref{MO-potential} can be expressed in terms of elliptic integrals by introducing the classical turning points $0<a<b$ through $V_{\rm quartic}(x)=\mu_F$ and $k=a/b$: 
\begin{align}
\label{eq:omell}
    & \omega = \frac{\pi\sqrt{2|g|}\,b}{2K(k)}, \qquad 
    \eta= \frac{2\omega}{\pi^2}K(k)\left((3-k^2)K(k)-4E(k)-\frac{\pi^2}{4K(k)}\right)\ .
 \end{align}
The small-$\mu$ expansions of these expressions were given earlier (see also appendix \ref{app:WKB-analytic}). We have also obtained analytic expressions for the long-string parameters $\omega(g)$ and $\eta(g)$ for 
the cubic and double-well potentials; they can be found in Appendix \ref{app:WKB-analytic}.

The crossover level $n_{\rm max}$ may be estimated by matching the Regge (short-string) trajectory to the long-string scale, equivalently by $\Delta_{\rm max}\simeq \eta_{\rm max}$. From the point of view of the 2D string dual, $n_\text{max}$ is the scale at which the tip of the fold reaches the value where the Liouville potential becomes significant. This occurs when 
\begin{equation}
    4 \pi \mu 
  e^{2\phi/\sqrt{\alpha'}}  \sim 1\ .
\end{equation}
Given the semiclassical relation \eqref{eq:foldedstring} between the maximum value of $\phi$ and $E_n \sim \sqrt{\frac{2n}{\alpha'}}$ for the cubic theory, this implies
\begin{equation}
    \frac{\phi}{\pi \sqrt{\alpha'}} \sim - \frac{1}{2 \pi }\log (4 \pi \mu) \sim  \sqrt{2 n^{\rm cubic}_{\rm max}},
\end{equation}
which agrees approximately with our formula for $ n^{\rm cubic}_{\rm max}$, see \eqref{nmaxcubic}. A similar argument gives the formula for $n_\text{max}$ for the quartic potential \eqref{eq:regge-phi4}.

\section{Discussion}

In this paper, we clarified the structure of adjoint states in the SU$(N)$ symmetric matrix quantum mechanics tuned near the critical coupling and proposed their interpretation in terms of 2D string theory. It is of obvious interest to study also the SU$(N)$ representations higher than the adjoint, whose dimensions grow faster than $N^2$. The allowed representations must contain an element with all weights equal to zero \cite{Gross:1990md}; their Young diagrams have the number of boxes divisible by $N$ \cite{Boulatov:1991xz, Hatsuda:2006xr}. The four representations whose dimensions grow as $N^4$ are the self-conjugate representations $B_2$ of dimension $\frac{1}{4} N^2 (N+3)(N-1)$ and $C_2$ of dimension $\frac{1}{4} N^2 (N-3)(N+1)$, as well as $A_2$ of dimension $\frac{1}{4}(N^2-4)(N^2-1)$ and its conjugate.
We find that the energy levels in the self-conjugate representations are given by sums of MO eigenvalues, $\Delta_i + \Delta_j$, up to $1/N$ corrections. The string theoretic interpretation of these states involves two folded strings, and higher representations involve multiple folds \cite{Balthazar:2018qdv}. A more detailed study of the representations higher than the adjoint, which are useful for the thermodynamics of the MQM, will be reported later.

In \cite{Gross:1990ub,Gross:1990md}, it was suggested that the contributions of non-singlet states to the free energy lead to a thermal phase transition. The non-singlets are suppressed at low temperatures, but they become important above the critical temperature, making this similar to the deconfinement transition in QCD. Evidence from the bootstrap for the existence of a critical temperature was reported in \cite{Cho:2024kxn}. From the point of view of the discretized string world sheet, this is expected to be the BKT transition caused by the vortices \cite{Gross:1990ub,Gross:1990md}. However, an estimate of the critical temperature in \cite{Gross:1990md} based on the WKB approximation appeared to produce a factor of $2$ difference from the continuum formula $T_{\rm BKT}=\frac{1}{4\pi\sqrt{\alpha'}}$. 
Due to our new finding that the WKB approximation \eqref{eq:WKB} is valid only for $n\gtrsim n_{\rm max}$, the equally spaced spectrum of the quartic theory begins at $\Delta\approx\frac{1}{\sqrt 2 \pi}\log \frac{4}{|g_c|\mu}$, which is twice the value assumed in \cite{Gross:1990md}. This appears to change the estimate of the critical temperature in MQM, but we leave a detailed study for the future.

We expect that the scattering phase of long strings \cite{Maldacena:2005hi, Fidkowski:2005ck, Balthazar:2018qdv} is conceptually related to the discrete ``bound states'' in our MO analysis via a L{\"u}scher-like relation. If one naively takes the approximation $V=-\tr\, X^2$, there is only the Liouville wall in the IR and we have a continuum of ``long string'' scattering states. But in the MQM that one actually gets from approaching $g\to g_c$ there is a ``UV wall'' in addition to the IR Liouville wall. So we have bound states instead of a continuum, but as $g\to g_c$ the continuum scattering phase shift can nevertheless be extracted from the theory in a finite box. Conversely, the scattering phase shift should allow us to smoothly interpolate between the low-energy Regge regime and the high-energy WKB approximation; such states have an energy $\Delta = \Delta_\text{max} + \mathcal{O}(1)$. Note that there are $\sim \log \mu$ states in this transition region. This number is large but is significantly smaller than the total number of Regge levels $\sim \log^2\mu$. We hope to report on the physics of this transition region in future work.

\subsection*{Supplemental material}
The data supporting the findings of this article are openly available in the Supplemental Material. This includes the scripts \texttt{solver.py}, \texttt{extrapolation.py}, and instructions \texttt{README.md} used to produce the numerical spectrum of Marchesini-Onofri equation in the case of quartic potential.

\section*{Acknowledgments}

We are grateful to Aleksandr Artemev, Minjae Cho, Barak Gabai, Alexey Litvinov, Juan Maldacena, Jessica Yeh, and Zechuan Zheng for useful discussions.
This work was supported in part by the Simons Foundation Grant No.~917464 (Simons Collaboration on Confinement and QCD Strings).
\appendix
\section{Analytic results for high-energy WKB\label{app:WKB-analytic}}
In this appendix, we collect the explicit expressions for high-energy WKB parameters $\omega(g)$ and $\eta(g)$ entering \eqref{eq:WKB} for the cubic and double-well potentials. We use the complete elliptic integrals of the first kind, $K(k)$, and second kind, $E(k)$, with their asymptotics near $k\to1$ 
\begin{equation}
\resizebox{\textwidth}{!}{$
    \begin{aligned}  
        K(k)=&\sum_{m=0}^{\infty}\left(\frac{(\frac{1}{2})_m}{m!}\right)^2(1-k^2)^{m}
        \left[\frac{1}{2}\log\frac{1}{1-k^2}+\psi(1+m)-\psi\left(\frac{1}{2}+m\right)\right]\ ,
        \\
        E(k)=&1+\frac{1}{2}\sum_{m=0}^{\infty}\frac{(\frac{1}{2})_m(\frac{3}{2})_m}{(m+1)!\,m!}(1-k^2)^{m+1}
        \left[\frac{1}{2}\log\frac{1}{1-k^2}+\psi(1+m)-\psi\left(\frac{1}{2}+m\right)-\frac{1}{(2m+1)(2m+2)}\right], \nonumber
    \end{aligned}
$}
\end{equation}
where $(n)_m$ is the Pochhammer symbol, $\psi(n)$ is the digamma function.

For the cubic potential, let $x_1<x_2<x_3$ be the three real turning points defined as the roots of $\mu_F=V_{\rm cubic}(x)$, and introduce the elliptic modulus $k_{\rm cubic}^2=\frac{x_2-x_1}{x_3-x_1}$. Then
\begin{equation}\label{cubicWKBparameters}
    \begin{aligned}
        \omega_{\rm cubic}(g) &= \frac{\pi\sqrt{2|g|(x_3-x_1)}}{2K(k_{\rm cubic})},\\
        \eta_{\rm cubic}(g) &=\, \frac{\omega_{\rm cubic}(g)}{\pi^2}K(k_{\rm cubic})\bigg[\frac{1+6gx_3}{g(x_3-x_1)}\,K(k_{\rm cubic})-6E(k_{\rm cubic})+\\
        &\quad  +\frac{\pi^2}{4g(x_2-x_1)K(k_{\rm cubic})}\left(\frac{x_2+3gx^2_2}{x_3-x_2}-\frac{x_1+3gx^2_1}{x_3-x_1}\right)\bigg]\ . 
    \end{aligned}
\end{equation}
Near criticality, $\mu\to0$, the corresponding asymptotics are
\begin{equation}\label{eq:cubic-WKB-near-crit}
    \begin{aligned}
        \omega_{\rm cubic}(\mu)
        &=\frac{2\pi}{\log\frac{8}{g_c^2\mu}}
        -\frac{15\pi g_c^2\mu}{\log\frac{8}{g_c^2\mu}}
        +\mathcal{O}\left(\frac{\mu}{\log^2\mu}\right),\\
        \eta_{\rm cubic}(\mu)
        &=\frac{1}{4\pi}\log\frac{8}{g_c^2\mu}
        -\frac{3}{\pi}
        -\frac{\pi}{\log\frac{8}{g_c^2\mu}}
        +\mathcal{O}\left(\mu\log\mu\right).
    \end{aligned}
\end{equation}

For the double-well potential, one must distinguish $\mu<0$ and $\mu>0$. When $\mu<0$ (so $\mu_F=-\mu>0$), the equation $\mu_F=V_{\rm dw}(x)$ has two real turning points $\pm a$, while the remaining roots are purely imaginary $\pm i b$ (with $a>b>0$); we define $ k_{\rm dw^-}^2=\frac{a^2}{b^2}\ge1$. In this case,
\begin{equation}\label{eq:dwminus-WKB}
    \begin{aligned}
        \omega_{\rm dw^-}(g)
        &=\frac{\pi}{2}\,\frac{\sqrt{2g}\,b}{K(i k_{\rm dw^-})},\\
        \eta_{\rm dw^-}(g)
        &=\frac{2\omega_{\rm dw^-}(g)}{\pi^2}\,K(i k_{\rm dw^-})
        \left[
        (3+k_{\rm dw^-}^2)K(i k_{\rm dw^-})-4E(i k_{\rm dw^-})
        -\frac{\pi^2}{4K(i k_{\rm dw^-})}
        \right].
    \end{aligned}
\end{equation}
For $\mu\to0^-$, we obtain
\begin{equation}\label{eq:dwm-WKB-near-crit}
    \begin{aligned}
        \omega_{\rm dw^-}(\mu)
        &=\frac{\pi}{\log\frac{4}{g_c|\mu|}}
        +\frac{3\pi g_c|\mu|}{\log\frac{4}{g_c|\mu|}}
        +\mathcal{O}\left(\frac{|\mu|}{\log^2|\mu|}\right)\ ,\\
        \eta_{\rm dw^-}(\mu)
        &=\frac{1}{2\pi}\log\frac{4}{g_c|\mu|}
        -\frac{4}{\pi}
        -\frac{\pi}{2\log\frac{4}{g_c|\mu|}}
        +\mathcal{O}\left(|\mu|\log|\mu|\right)\ ,
    \end{aligned}
\end{equation}
where we also used the identity $K(ik)=\frac{1}{\sqrt{1+k^2}}K\left(\frac{k}{\sqrt{1+k^2}}\right)$. 

When $\mu>0$, all four turning points $\pm a,\pm b$ are real and can be ordered as $0<a<b$; we introduce $k_{\rm dw^+}^2=1-\frac{a^2}{b^2}\in(0,1)$. Then
\begin{equation}\label{eq:dwplus-WKB}
    \begin{aligned}
        \omega_{\rm dw^+}(g)
        &=\frac{\pi\sqrt{2g}\,b}{K(k_{\rm dw^+})}\ ,\\
        \eta_{\rm dw^+}(g)
        &=\frac{\omega_{\rm dw^+}(g)}{2\pi^2}\,K(k_{\rm dw^+})
        \left[
        (2-k_{\rm dw^+}^2)K(k_{\rm dw^+})-4E(k_{\rm dw^+})
        -\frac{3\pi^2}{2K(k_{\rm dw^+})}
        \right]\ .
    \end{aligned}
\end{equation}
For $\mu\to0^+$, this yields
\begin{equation}\label{eq:dwplus-exp}
    \begin{aligned}
        \omega_{\rm dw^+}(\mu)
        &=\frac{2\pi}{\log\frac{4}{g_c\mu}}
        -\frac{6\pi g_c\mu}{\log\frac{4}{g_c\mu}}
        +\mathcal{O}\left(\frac{\mu}{\log^2\mu}\right)\ ,\\
        \eta_{\rm dw^+}(\mu)
        &=\frac{1}{4\pi}\log\frac{4}{g_c\mu}
        -\frac{2}{\pi}
        -\frac{3\pi}{2\log\frac{4}{g_c\mu}}
        +\mathcal{O}(\mu\log\mu)\ .
    \end{aligned}
\end{equation}
For completeness, we provide more precise expansions for the quartic case
\begin{equation}\label{eq:quartic-WKB-near-crit}
    \begin{aligned}
        \omega_{\rm quartic}(\mu)
        &=\frac{\pi \sqrt2}{\log \frac{4}{|g_c|\mu}}
        -\frac{3\pi\sqrt{2}|g_c|\mu}{\log\frac{4}{|g_c|\mu}}
        +\mathcal{O}\left(\frac{\mu}{\log^2\mu}\right),\\
        \eta_{\rm quartic}(\mu)
        &=\frac{1}{2\sqrt2\pi}\log\frac{4}{|g_c|\mu}
        -\frac{2\sqrt{2}}{\pi}-\frac{\pi}{\sqrt{2}\log\frac{4}{|g_c|\mu}}
        +\mathcal{O}(\mu\log\mu)\ .
    \end{aligned}
\end{equation}
The corresponding crossover levels $n_{\rm max}$, defined by $\Delta_{\rm max}\simeq\eta_{\rm max}$, are
\begin{align}
    \label{eq:nmax-phi4}
    n^{\rm quartic}_\text{max}&\approx\frac{1}{4\pi^2}\log^2\frac{4}{e^2|g_c|\mu} -\frac{1}{3}\ ,\\
    \label{eq:nmax-phi3}
    n^{\rm cubic}_{\rm max}&\approx\frac{1}{8\pi^2}\log^2\frac{8}{e^2g^2_c\mu}-\frac{5}{12}\ ,\\
    \label{eq:nmax-dw-negative}
    n^{\rm dw^-}_{\rm max}&\approx
    \frac{1}{2\pi^2}\log^2\frac{4}{g_c|\mu|}-\frac{1}{3}\ ,\\
    \label{eq:nmax-dw-positive}
    n^{\rm dw^+}_{\rm max}&\approx
    \frac{1}{8\pi^2}\log^2\frac{4}{e^2g_c\mu}-\frac{1}{6}\ .
\end{align}

\section{Numerical diagonalization}\label{app:numerics}
We start with the MO equation \eqref{MO-eq} for the quartic potential \eqref{eq:quartpot}. We rescale the $x \to  a x$ coordinate and introduce $k^2 =-\frac{g a^4}{\mu_F}=\frac{a^2}{b^2}$ such that 
\begin{equation}
    \Delta\,\Phi(x)= \frac{1}{\sqrt{1+k^2}}{\fint_{-1}^{1}\limits \frac{\d y}{\pi}\frac{\sqrt{(1-y^2)(1-k^2y^2)}}{(x-y)^2}} \left( \Phi(x) - \Phi(y) \right)\ .
\end{equation}
Here $0<k^2\leq1$ for $g<0$, while $-1<k^2<0$ for $g>0$, since $b$ becomes purely imaginary. We discretize the integrand on a grid
\begin{equation}
    x_m=\cos\phi_m\ , \quad\phi_m=\frac{m\pi}{M+1}\ , \quad m = 1, 2, \dots, M\ .
\end{equation}
Then the MO equation is
\begin{align}
    \Delta_n\,\Phi_n(x_j)&=\sum_{m=1}^{M}K_{jm}\Phi_n(x_m)\ ,\\
    K_{jm}&=-\frac{1}{M+1}\frac{\sin^2\phi_m}{\sqrt{1+k^2}}\frac{\sqrt{1-k^2\cos^2\phi_m}}{(\cos\phi_j-\cos\phi_m)^2}\ , \qquad j\ne m\ ,\\
    K_{jj}&=-\sum_{m \ne j} K_{jm}\ .
\end{align}
For numerical stability, it is more convenient to conjugate this operator $K \to H=DKD^{-1}$, where $H$ is a Hermitian operator and $D$ is the diagonal matrix $D_{jm}=\delta_{jm}{\sin\phi_m(1-k^2\cos^2\phi_m)^{1/4}}$.
\begin{equation}
    H_{jm}=-\frac{1}{M+1}\frac{\sin\phi_j\sin\phi_m}{\sqrt{1+k^2}}\frac{(1-k^2\cos^2\phi_j)^{1/4}(1-k^2\cos^2\phi_m)^{1/4}}{(\cos\phi_j-\cos\phi_m)^2}\ , \quad j \ne m\ ,
\end{equation}
The low-lying spectrum can be efficiently computed using the Lanczos algorithm.
We then performed a large $M$ extrapolation for each eigenvalue that was obtained, e.g., we fit a function
\begin{equation}
    \Delta^{M}_n = \Delta_n + \frac{1}{M} a_n + \frac{1}{M^2} b_n + \frac{1}{M^3} c_n + \frac{1}{M^4} d_n\ .
\end{equation}
The results of this large $M$ extrapolation are reported in Figure \ref{fig:spectrum}. For this plot we used values of $M$ ranging from $M= \{25000, 26000,27000, \dots, 40000\}$.
Note that the eigenfunctions are highly oscillatory for large energies and therefore, to probe the ``long string'' states, one has to consider rather large values of $M$ to accurately find these states for small $\mu$. 

For the cubic potential, we used the same method but with $M = 40000$ to 46000 with increments of 2000. We performed a large $M$ extrapolation using the ansatz $\Delta_n^M=\Delta_n+a_n/M+b_n/M^2$. We used these slightly different parameters since at $\mu = 10^{-14}$ the value of $n_\text{max}$ is appreciably lower in the cubic than in the quartic; hence the eigenvalues at $n\gtrsim 36$ for the cubic are not well-converged for $M \lesssim 4 \times 10^4$.

In Table \ref{tab:numerics} we report the first $5$ eigenvalues for both the cubic and quartic at $\mu=0$; we see excellent agreement with the analytic predictions even at small values $n$.
\begin{table}[ht]
    \centering
    \begin{tabular}{|cIc|c|cIc|c|c|}
    \hline
     & \multicolumn{3}{cI}{Quartic} & \multicolumn{3}{c|}{Cubic}\\
    \noalign{\hrule height 1.2pt}
    $n$ & $\Delta^{\rm numeric}_n$ & $\Delta^{\rm analytic}_n$ & $\delta_n$  & $\Delta^{\rm numeric}_n$  & $\Delta^{\rm analytic}_n$ & $\delta_n$
    \\
    \hline
    1 & 0.7415737 & 0.7326768 &$ 1.2 \times 10^{-2} $ & 0.7297836  & 0.7283212 & $2.0 \times 10^{-3}$ \\
    \hline
    2 & 1.2612196 & 1.2599306 & $1.0 \times 10^{-3}$  & 1.2436438  & 1.2435547 & $7.2\times 10^{-5}$ \\
    \hline
    3 & 1.6819242 & 1.6816726 & $1.5\times 10^{-4}$  & 1.6591433  & 1.6591349 & $5.1\times 10^{-6}$ \\
    \hline
    4 & 2.0436636 & 2.0436040 & $2.9\times 10^{-5}$  & 2.0171638  & 2.0171628 & $5.2\times 10^{-7}$\\
    \hline
    5 & 2.3656862 & 2.3656700 & $6.8\times 10^{-6}$  & 2.3364734 & 2.3364733 & $7.1\times 10^{-8}$ \\
    \hline
    \end{tabular}
    \caption{$\Delta_n$ for first $5$ levels from the numerical method (large $M$ extrapolation) for quartic and cubic theories. We also show the analytic prediction $\Delta^\text{analytic}$ from \eqref{eq:regge-phi4}, \eqref{eq:regge-phi3} and the relative error $\delta_n\equiv {|\Delta^{\rm numeric}_{n}-\Delta^{\rm analytic}_{n}|}/{\Delta^{\rm numeric}_{n}}$. 
    }
    \label{tab:numerics}
\end{table}
\section{Effect of the linear dilaton background}\label{app:Dilaton-back}
Let us consider the world sheet action, which includes the linear dilaton background and the Liouville potential,
\begin{equation}
    S\left[\phi\right]=\int\d^2 \sigma  \sqrt{-\operatorname{det} g}\left[
    \frac{1}{4 \pi \alpha^{\prime}}
    g^{a b} (-\partial_a X^0 \partial_b X^0 +\partial_a \phi \partial_b \phi) + \frac{1}{4 \pi} Q \phi R(g)- \mu e^{2\phi/\sqrt{\alpha'}}\right]\ .
\end{equation}
With these conventions, $c_\phi= 1+ 6\alpha' Q^2$. In the special case $c_\phi=25$, where $Q=\frac{2}{\sqrt{\alpha'}}$, it is more appropriate to take the Liouville potential as $-\mu\frac{\phi}{\sqrt\alpha'}e^{2 \phi/\sqrt{\alpha'}}$ \cite{Polchinski:1990mf} (see also \cite{Seiberg:1990eb, Klebanov:1991qa}). For a recent discussion of this issue from the probabilistic Liouville theory perspective, see \cite{Duplantier:2012saa, Chatterjee:2024phq}.

We work in the static gauge where $X^0 = \tau$ and set $g=g_c$ so that we focus on the dynamics far from the Liouville wall. A classical solution \cite{Maldacena:2005hi} is 
\begin{equation}
    \phi(\tau,\sigma)
    = \phi_0' - Q \alpha' \log \left( \cosh \frac{\tau}{\alpha' Q} + \cosh \frac{\sigma}{\alpha' Q} \right)\ .
\end{equation}
So the tip $\phi_\text{tip} = \phi(\tau,0) $ at $\sigma = 0$ moves according to
\begin{equation}
    \phi(\tau) = \phi_0 - 2 Q \alpha^{\prime}  \log \cosh (\frac{\tau}{2\alpha' Q})\ ,
\end{equation}
where $\phi_0=\phi'_0-Q\alpha'\log{2}$.
From this solution we can derive the space-time energy, which in static gauge is just $E = \frac{1}{2\pi \alpha'}\int \d \sigma $. This quantity is divergent; to regulate it we cut off the world sheet at $\phi = \phi_c$. Then the spacetime energy for the folded string is
\begin{equation}
    E = \frac{1}{\pi \alpha'}(\phi_0 - \phi_c)
    = \frac{\phi-\phi_c}{\pi \alpha^{\prime}}-\frac{Q}{\pi } \log (1-v^2)\ .
\end{equation}
Here $ v= \d \phi/ \d \tau $.
We can rewrite the Hamiltonian in canonical variables, by using the relation $v = \dot{\phi} = \partial H/\partial P$ or $\d P = \frac{1}{v} \d E$:
\begin{equation}
    P = \int \frac{1}{v} \frac{\partial E}{\partial v} \,\d v = \frac{2Q}{\pi }\int \frac{\d v}{1-v^2}  = \frac{2Q}{\pi} \text{ arctanh\,} v\ .
\end{equation}
 Thus $1 - v^2 = 1/\cosh^2\left( \frac{\pi}{2Q} P \right)$, or 
\begin{equation}
    E =  \frac{\phi-\phi_c}{\pi \alpha'} + 
     \frac{2Q}{\pi} \log \cosh \left(\frac{\pi}{2Q} P \right)\ .
\end{equation}
Expanding at small $P$, we get a term
\begin{equation}
    E \supset \frac{1}{2M}P^2 , \quad M = \frac{2 Q}{\pi }\ .
\end{equation}
\bibliography{MyBib}
\bibliographystyle{MyStyle}

\end{document}